\documentclass[12pt,authoryear]{elsarticle}

\usepackage{hyperref}
\usepackage{graphicx} 

\journal{Planetary and Space Science}

\begin{document}

\begin{frontmatter}

\title{Surface charging of thick porous water ice layers relevant for ion sputtering experiments}

\author[label1]{A. Galli\fnref{myfootnote}} \author[label1]{A. Vorburger} \author[label1]{A. Pommerol}
\author[label1]{P. Wurz}  \author[label1]{B. Jost} \author[label1]{O. Poch} \author[label1]{Y. Brouet} 
\author[label1]{M. Tulej} \author[label1]{N. Thomas}

\fntext[myfootnote]{Corresponding author, E-mail address: \textit{andre.galli@space.unibe.ch}}

\address[label1]{Physikalisches Institut, University of Bern, Bern, Switzerland}

\begin{abstract}
We use a laboratory facility to study the sputtering properties of centimeter-thick
 porous water ice subjected to the bombardment of ions and electrons to better understand the formation of 
exospheres of the icy moons of Jupiter. Our ice samples are as similar as possible
 to the expected moon surfaces but surface charging of the samples during 
ion irradiation may distort the experimental results. We therefore monitor 
the time scales for charging and discharging of the samples when subjected to a beam of ions. 
These experiments allow us to derive an electric conductivity of deep porous ice layers. 
The results imply that electron irradiation
and sputtering play a non-negligible role for certain plasma conditions at the icy moons of Jupiter.
The observed ion sputtering yields from our ice samples are similar to previous experiments where compact ice 
films were sputtered off a micro-balance. 
\end{abstract}


\end{frontmatter}

\section{Introduction}\label{sec:introduction}

The vast majority of celestial objects in the outer Solar System are bodies with ice-rich surfaces. 
This includes icy moons like Europa, 
Ganymede, Callisto, Enceladus, Triton, but also the Trans-Neptunian Objects and comets. The surface of these objects directly 
interacts with the space environment, resulting in sputtering, radiolysis, and sublimation. These processes lead to a tenuous, 
surface-bound atmosphere. For the Galilean icy moons, it is often referred to as `exosphere' since the mean free path 
length usually is larger than the atmospheric scale height \citep{mar07,pla12}. 
A few species of these atmospheres have been measured by remote sensing techniques.
At Callisto, only CO$_{2}$ has been detected so far, with a surface pressure as low as 10$^{-8}$ mbar \citep{car99}. 
For Ganymede and Europa, \citet{hal98} reported the detection of oxygen airglow, 
implying atmospheric O$_{2}$ pressures on the order of 10$^{-8}$ mbar (column densities of several 10$^{14}$ cm$^{-2}$). 
Atomic oxygen, one to two orders of magnitude more tenuous, was observed by \citet{hal95} at Europa. 
The O and O$_{2}$ column densities of Europa were later confirmed during the CASSINI flyby with the 
Ultraviolet Imaging Spectrograph \citep{han05}. 
%Recently, however, a debate regarding these UV data 
%was triggered by \citet{she14}. They interpret the data as to indicate a much more tenuous atmosphere 
%dominated by atomic oxygen with column densities of the order of 10$^{12}$ cm$^{-2}$. \textit{See rebuttal
%paper by McGrath.}
A tenuous Na and K atmosphere around Europa was detected with ground-based observation 
in the visible spectral range \citep{bro96,bro01}. 
\citet{bar97} reported the detection of atomic hydrogen in Ganymede's atmosphere with the Galileo
ultraviolet spectrometer, whereas \citet{spe95} and \citet{nol96} reported the detection of condensed
or trapped O$_{2}$ and O$_{3}$ in the surface of Ganymede.
The latter discovery showed that the atmospheric oxygen species are indeed surface-bound.

Unfortunately, properties of ices at low pressures and temperatures, such as the 
sputtering efficiency due to ion bombardment, are difficult to predict theoretically and 
it is not always clear how results from laboratory experiments should be applied to real ice surfaces
in the Solar System \citep{joh04}. 
This limits the predictive capability of any surface and atmosphere model. 
The JUICE mission \citep{ESA14}, scheduled to visit Europa, Ganymede, and Callisto in the years 2030--2032, 
will allow to directly 
sample the particles ejected from the surface and also to observe the atmosphere at infrared, visible, and 
ultraviolet wavelengths. The design of instruments would benefit from better constrained 
parameters for surface release processes.

Over the last two decades, several laboratory experiments simulating particle irradiation of water ice on 
the surface of icy moons have been performed
\citep{shi95,orl03,fam08,shi12}, and these data have been put into the context of planetary science
\citep{joh04,cas10}. Most approaches aimed at deriving the sputtering yields,
i.e., the ratio of ejected water ice molecules per impinging ion or electron as a function of impactor energy. 
The favourite technique used so far consists of vapour depositing a thin film ($100-1000$ nm) 
of compact (density $\approx 0.9$ g cm$^{-3}$) amorphous ice onto a quartz microbalance \citep{fam08,shi12}.
The ice film is then gradually sputtered and the observed frequency change of the quartz crystal allows to deduce 
the sputtering yield. 
In contrast, we experiment with ions impacting a 0.9 cm thick sample of porous water ice to simulate
the regolith on the surface of an icy moon. ``Porosity'' in this study always implies macroporosity, i.e., 
micron-sized gaps between grains, as opposed to  nano-scale pores within a grain \citep{rod11}.
Our approach allows us to produce samples with chemical impurities
and to study the impact of porosity on release processes. So far, sputtering models
disagree whether the sputtering yield from 
a porous solid should be substantially lower \citep{joh89,cas05} or similar \citep{cas13} 
compared to the same non-porous solid.

Experimental difficulties arising from sputtering experiments with a cm-thick ice sample
 include a strong surface charging due to ion irradation and a more urgent need to characterize the
sample (for instance density, chemical alterations, and possible temperature gradients between the bottom
and the surface). Our first sputtering experiments with H$^{+}$, O$^{+}$, O$_2^{+}$, and Ar$^{+}$ ions \citep{gal15}
 indicated that surface charging of the ice sample could affect estimates of the sputtering yield. 
The charging also limited the number of unbiased 
experiments per day since the discharging times exceeded one hour for cold ($T<100$ K) ice if no countermeasures were employed. 
Our improved set-up for sputtering experiments with porous ice allows us to detect and counteract 
surface charging caused by ion irradiation. This paper describes the new set-up and
presents the first Ar$^{+}$ sputtering results obtained with it. 

\section{Experiment set-up}\label{sec:experiment}

The MEFISTO test facility for space instrument calibration has been developed over more than a decade \citep{mar01,hoh02,hoh05}.
It consists of a vacuum chamber and an electron-cyclotron-resonance ion source.
\citet{wie16} examined the charge exchange of an ion beam on an ice surface in MEFISTO
to prepare for an energetic neutral atom imaging instrument on board JUICE.
We present here how the same test facility and a similar set-up can be used to simulate ion 
sputtering of porous icy material in space.

\subsection{Sample preparation}

\begin{figure}
\begin{center}
\includegraphics[width=1.0\textwidth]{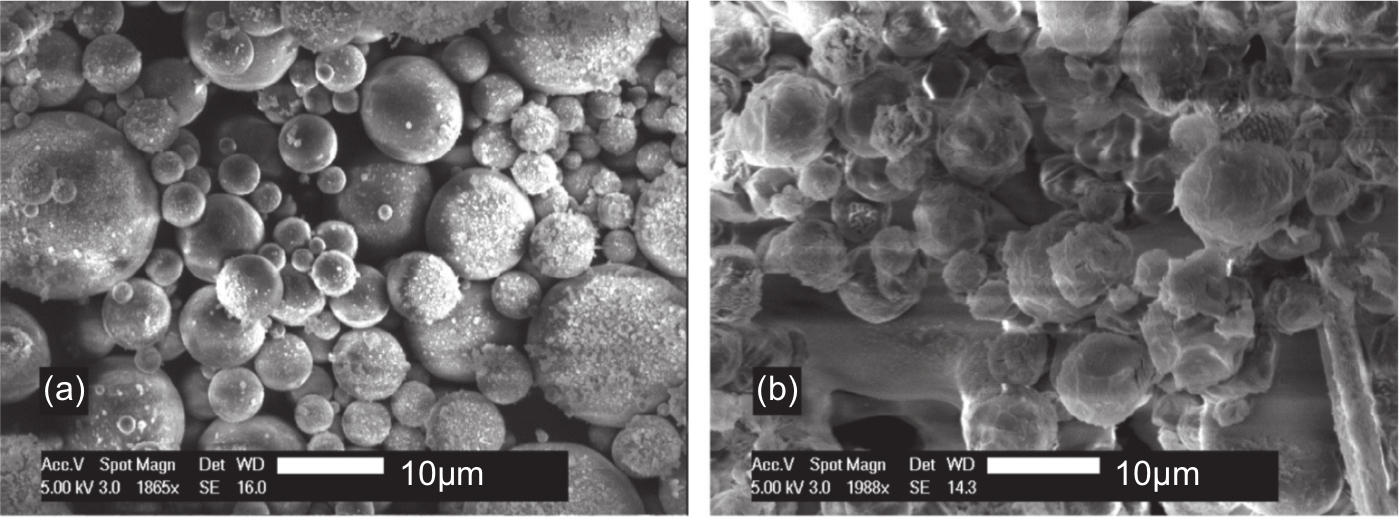}
\end{center}
\caption{Scanning electron microscope images of two samples of water ice,
reproduced from Fig. 5 in \citet{jos16}. The spatial scale is shown by the scale bars to the bottom. 
a) Fresh sample produced by directly spraying the water vapour on a cold plate. 
 The shape of the particles is almost perfectly spherical. Some of the particles
are covered with frost, which had condensed on the sample during the transfer to the microscope.
b) Fresh sample produced by mixing the fresh ice with liquid nitrogen. The shape looks more irregular
because the suspension was stirred up with a spoon before it was poured into the sample holder. 
 \label{fig:freshice}}
\end{figure}

We prepared porous water ice with the SPIPA (Setup for the 
Production of Icy Planetary Analogs) setup described
as Method \#1 by \citet{jos16}. 
An ultrasonic nebulizer produces micrometer sized droplets of salty water 
(99\% weight percent de-ionized water and 1\%~NaCl). 
We added the NaCl to discriminate sputtering of the salty ice 
from sputtering of the top frost layer originating from exposure to air. 
The water droplets were frozen and piled up in a cooled bowl ($T=-50^{\circ}$C) as spherical ice particles
with diameters of $4.5\pm2.5$ $\mu$m (left panel in Fig.~\ref{fig:freshice}).
Once enough ice was produced, we poured liquid nitrogen into the bowl and stirred up the suspension with a
spoon. This method prevented the grains from sintering and allowed
us to pour the ice into the sample holder.
A microscope image of ice grains after this step is shown in the right panel of Fig.~\ref{fig:freshice}. 
We then compressed the ice in the sample holder with the spoon 
until we had a smooth surface at the same level as the metal rim.
We did this to improve the temperature conductivity compared to the
very porous ice in the ice bowl. The bulk density of the final ice
sample ranged between 200 and 300 kg m$^{-3}$, corresponding to an average porosity of 75\%.

The ice in the sample holder is a realistic approximation for surfaces on the icy moons of Jupiter, 
considering the limited knowledge about these surfaces. 
The temperature of the ice in the sample holder was kept at temperatures between 90 and 140 K,
representative for the surfaces of Ganymede \citep{mar07} and Europa \citep{rod09}.
The Na content of our ice is representative for ice-rich areas on Europa's surface
(the average mass abundance of Na on Europa's ridged plains calculates to 2\% \citep{shir10}).
\citet{dom97} assume porosities ranging 
between 0.4 to almost 1.0 for the Galilean moons to fit solar phase curve reflectances. The authors 
conclude with respect to Europa: ``The high porosity
for Europa is consistent with a medium to fine-grained water frost.'' The grain size on the surface
of the Galilean moons probably is somewhat larger than in our unsintered ice. \citet{cal95}
use typically 100 $\mu$m sized ice grains to fit infrared spectra.        
When the ion sputtering experiment in the vacuum chamber starts, the ice grains will have undergone 
sintering to a small degree because of the residence time in the freezer and in the vacuum chamber
before the latter is cooled down to experiment temperatures. Moreover, the sample is covered
with a thin water frost layer. However, sintering and frost deposition are also considered 
to reproduce infrared spectra of icy satellites \citep{gru99}. 

\subsection{Experiment procedure}\label{sec:procedure}

\begin{figure}
\begin{center}
\includegraphics[width=1.0\textwidth]{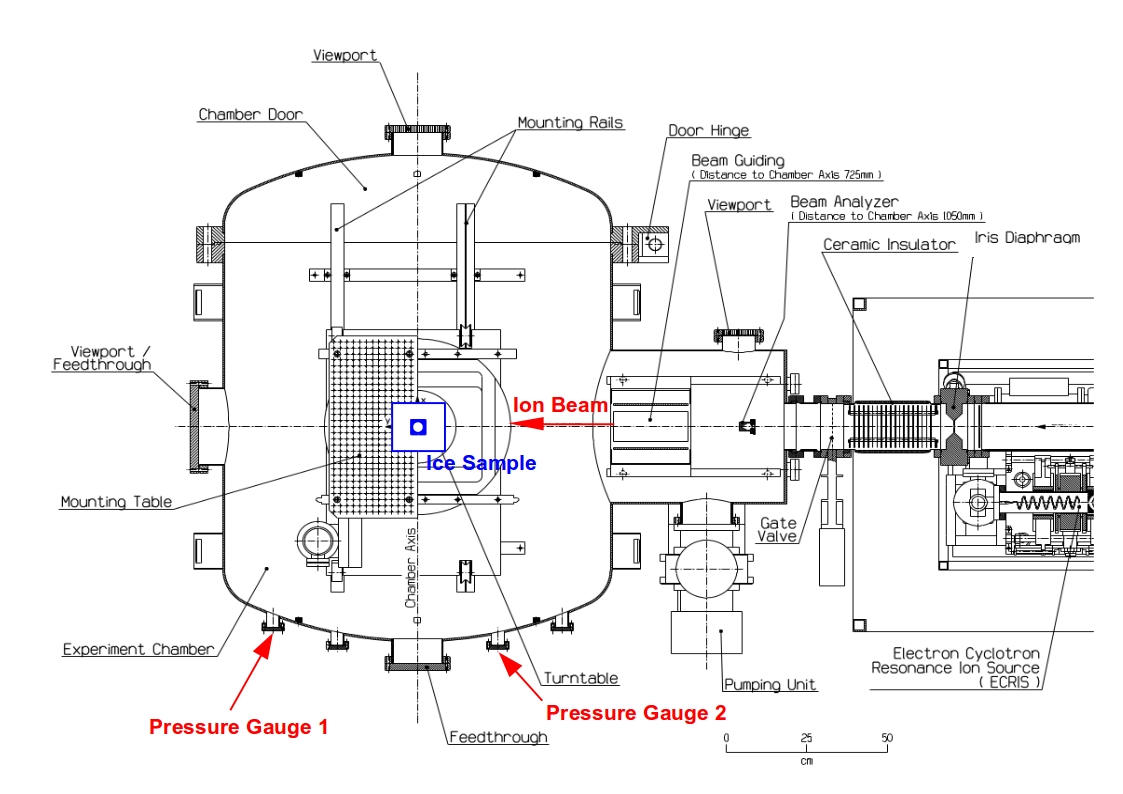}
\end{center}
\caption{Experiment set-up in the MEFISTO vacuum chamber. The drawing shows the ion source, 
the vacuum chamber, the ice sample on the cooling plate, and the pressure gauges. We selected
an incidence angle of $30^{\circ}$ between the ice sample and the ion beam during the present study.
In addition, a quadrupole mass spectrometer and an electron gun are mounted in the chamber 
roof 80 cm above the ice sample.}\label{fig:schematic}
\end{figure}

\begin{figure}
\begin{center}
\includegraphics[width=1.0\textwidth]{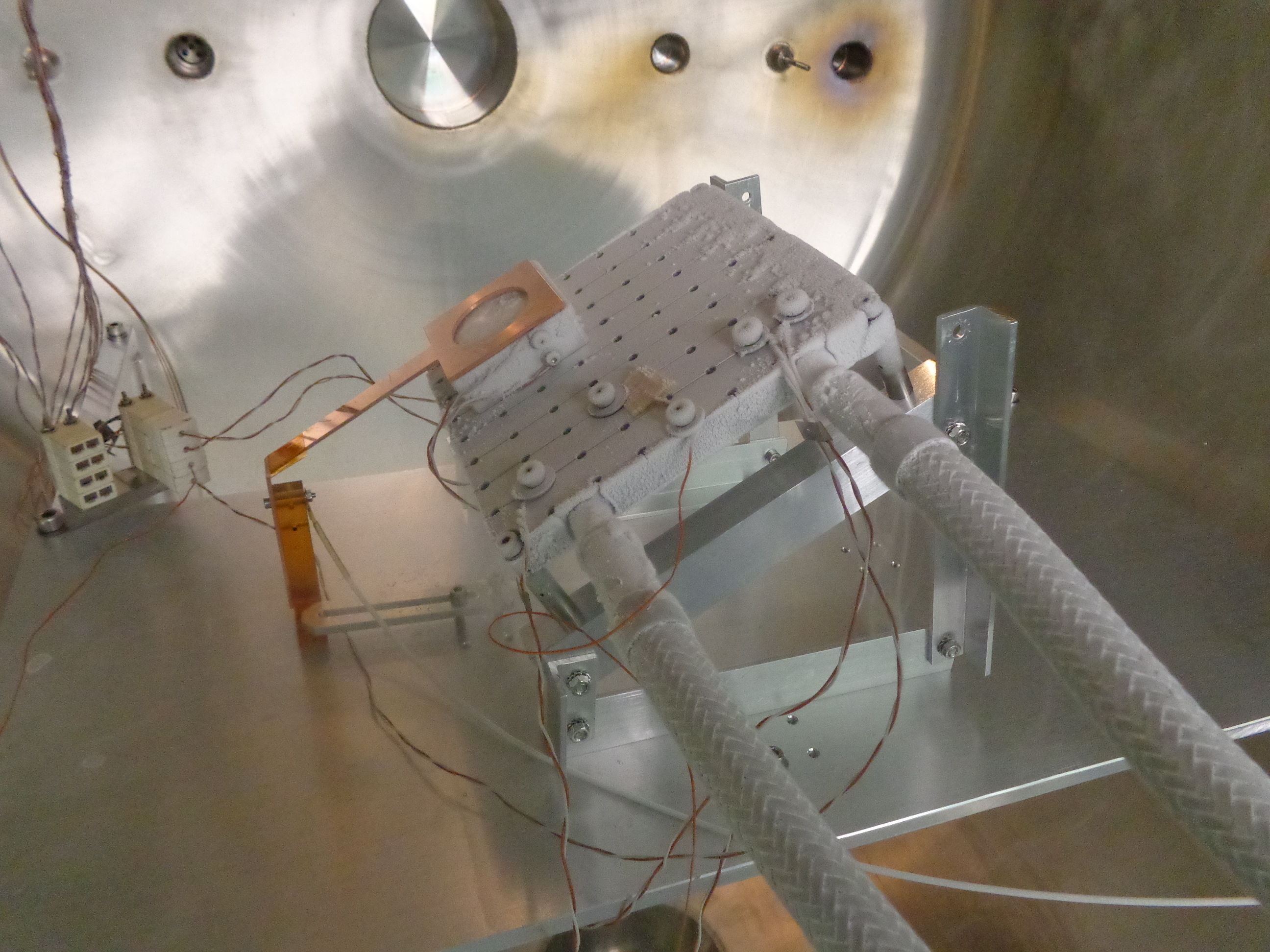}
\end{center}
\caption{Photograph of the experiment set-up at the beginning of the experiment. The ice has been
inserted in the sample holder and the copper ring has been positioned above the ice sample holder
so that it covers its rim but not the circular centre with the 0.9 cm deep ice layer.}\label{fig:picture}
\end{figure}

Figures \ref{fig:schematic} and \ref{fig:picture} show a drawing and a photograph of our experiment set-up.
The porous ice is inside an aluminum sample holder with a circular mound (radius 2 cm and depth 0.9 cm).
The sample holder itself is screwed onto a steel plate which is being cooled with liquid nitrogen throughout 
the entire experiment. The ion beam impacts
the ice sample at a flat angle of $30^{\circ}$ (from the left hand side in Fig.~\ref{fig:picture}). 
This angle could not be varied
during the suite of experiments described here as we tested a simplified set-up without a maneuverable mounting table.
We will vary the ion incidence angle in future experiments because the sputtering yield is expected 
to decrease at steeper angles (see Section \ref{sec:theory}). 
Electrostatic deflection plates (labelled ``Beam Guiding''
in the sketch in Fig.~\ref{fig:schematic}) between the beam exit
and the ice serve to direct the ion beam to any target area on the ice, the copper ring, or on
the cooling plate around it. The copper ring around the ice acts as a Faraday cup measuring
the current of impacting ions and electrons. The Faraday cup is operated at a positive potential of 18 V to prevent
escaping secondary electrons from distorting the measurement.
An electron gun (nominal energy range 100 eV to 10 keV, manufacturer: Kimball Physics) 
and a mass spectrometer (HAL quadrupole gas analyser, manufacturer: Hiden Analytical) are mounted to the ceiling of the
vacuum chamber, 80 cm above the ice sample. The electron gun is necessary to counter the surface charging effect due to ions. 
The photograph in Fig.~\ref{fig:picture} was obtained at the start of the experiment 
when water frost from the ambient air had condensed
on all cold surfaces. Afterwards we closed the chamber and started fast evacuation. A close-up image of the ice sample
obtained with a webcam was saved in a regular interval.
Because the electron gun, the mass spectrometer, and the ion beam require a vacuum pressure of $10^{-6}$ mbar or lower
we started sputtering experiments only after pumping and cooling for roughly 20 hours. The experiment
usually ran during one week. 

Throughout the measurements we relied on temperature sensors to monitor the temperature
of the chamber wall, the cooling plate, and of the ice sample holder (note the wires attached
to screws in Fig.~\ref{fig:picture}). The surface temperature of the ice
cannot be measured with a sensor as the latter would introduce a heat source to the porous ice and
commercial infrared radiometers are usually not sensitive to surface temperatures below 170 K \citep{bro08}. 
Nevertheless, the effective
surface temperature during sputtering experiments must have been close to the temperature
measured on the sample holder for the following reason:
The saturation vapour pressure for water frost at 153 K is $1.4\times10^{-7}$ mbar \citep{and07}.
The chamber pressure reached this level after 20 hours and the measured temperature was 143 K. We did not observe sublimation
of the ice in the sample holder during 8.5 hours at these conditions.  
The true surface temperature of the ice therefore must have matched the measured temperature of the 
sample holder to better than 10 K.

Due to surface charging, an ion beam that is initially directed at the centre of the
ice is deflected sideways until it hits the copper ring. We either waited for the 
ice in the sample holder to discharge on its own or we switched on the electron gun to neutralize the charged surface.
We continuously monitored the temperature sensors, the chamber pressure, 
and the current on the Faraday cup above the ice. From the timeseries of the current
we derived the charging and discharging timescales of the ice sample at different temperatures.
Two different pressure gauges at different places were used to measure the chamber pressure.  
Pressure Gauge 1 in Fig.~\ref{fig:schematic} is a Stabil-Ion pressure gauge measuring at an
accuracy of 10\% and at a time resolution of 1 s, whereas Pressure Gauge 2 is a
less precise baffled cold cathode gauge. %, i.e., the sensor is situated in a tube behind a disc. 
Sputtering of the water ice leads to a pressure rise in the vacuum chamber.
To detect the small pressure increase due to ion sputtering, only the Stabil-Ion pressure 
gauge was found to be precise enough. The pressure rise due to electron irradiation 
could be detected by both pressure gauges because
the electron flux was orders of magnitude higher than the ion flux.  

\section{Theory}\label{sec:theory}

\subsection{Surface charging and discharging}\label{sec:theory_charging}

Since water ice is an excellent insulator, irradiation with positively charged ions will cause a positive surface charge
in spite of the added NaCl salt. 
If the electric field created by the implanted ions on the ice surface is homogeneous,
the ice in the sample holder can be considered a parallel plate 
capacitor and the charging time $t_c$ calculates to
\begin{equation}
t_c = \frac{U}{i}\frac{ \varepsilon \varepsilon_0 A}{d}
\label{eq:fullycharged}
\end{equation}
In the above equation, $i$ is the ion current, $\varepsilon_0 = 8.85 \times 10^{-12}$ F m$^{-1}$ is the vacuum permittivity,
$\varepsilon$ is the relative permittivity of cold porous ice,
$d$ = 0.009 m is the thickness of the ice layer, and $A$ = $\pi r^2$ = 0.0013 m$^{2}$ is the ice surface area inside the sample holder. 
Upon irradiation with ions, the ice surface will charge up to an average surface potential $U$
that will deflect any additional ion onto the copper ring around the ice. 
If we again use the parallel plate capacitor as a model, this 
surface potential $U$ is related to the acceleration voltage of the ions $U_i$ via
\begin{equation}
\frac{U}{d} = \frac{\sin(\alpha)\cos(\alpha)U_i}{r}
\label{eq:effective_potential}
\end{equation} 
with $\alpha=30^{\circ}$ the incidence angle between the ion beam and the ice plane.
The other unknown in Equation \ref{eq:fullycharged} is $\varepsilon$.
In this study, we will only consider the real part of $\varepsilon$,
since its imaginary part is orders of magnitude lower \citep{mae06}.
\citet{shi12} adopted the value of $\varepsilon=3.0$ from \citet{wan08} for compact water ice.
A similar value of $\varepsilon=3.1$ is recommended by \citet{mae06} for ordinary water ice (Ih) below $T=240$ K.
The permittivity of water ice hardly varies with temperature for the range of 90 to 140 K considered
in our study \citep{mae06}.
The density of our ice samples closely resembles dry snow. We will therefore 
rely on \citet{sug10} who measured $\varepsilon=1.5$ at microwave 
frequencies for dry snow with a density of 300 kg m$^{-3}$. This agrees
with the value given by \citet{bro16} for pure porous water ice with 300 kg m$^{-3}$ bulk density. 
Equations~\ref{eq:fullycharged} and ~\ref{eq:effective_potential} then yield $t_c \approx 4$ s and a surface potential of 
$U\approx 2$ kV for an ion beam with $i = 1$ nA and an energy $U_ie^{-}=10$ keV.

\citet{shi10,shi12} directly measured the surface potential of amorphous water ice films 
with a Kelvin probe. 
They vapour deposited ice films with hundreds to thousands of nm thickness onto the tip of 
a microbalance cooled to temperatures between 20 K and 160 K.
They found that the surface potential increased asymptotically to an equilibrium during irradiation with an Ar$^{+}$ beam.
The timescale to charge the ice surface was hundreds of seconds.
The formulae describing the charging of the ice derived by \citet{shi10,shi12} 
cannot be directly applied to our experiments because of the different types of samples
and the completely different dimensions.  
In particular, the ice surface potential derived by \citet{shi10} contains the distance
between the trapped charges, i.e., ionization depth, and the metal substrate
underneath the ice. The formula describes well the charging of ice layers whose thickness
has the same order of magnitude as the ionization depth
(200 nm for 100 keV Ar$^{+}$). In our case, however, this distance would be on the order
of 0.9 cm and the asymptotic potential would be much larger than the corresponding energy of the ions.
In reality, the surface potential will increase to the fraction of the ion acceleration voltage sufficient to deflect
subsequent ions away from the ice onto the surrounding metal surfaces.  

When the ion bombardment is stopped, the surface discharges in the most simple case with a decay constant $\lambda_d$,
\begin{equation}
Q(t) =Q_{0} \exp(-\lambda_d t)
\end{equation}
If the ice surface is interpreted again as a parallel plate capacitor (see Eq.~\ref{eq:fullycharged}),
the observed time constant for discharging is related to the static conductivity $\kappa$ of the porous ice via
\begin{equation}
\kappa =\varepsilon \varepsilon_0 \lambda_d
\label{eq:kappa}
\end{equation}

\citet{pet93} performed a meta-study of previous conductivity measurements for ordinary water ice (Ih).
The conductivity depends on temperature, $\kappa\sim \exp(-E_a/k_BT)$,
with the activation energy $E_a$ ranging from 0 to 0.7 eV. The value is dominated by the chemical impurities 
(Cl$^{-}$ concentrations for instance) inside the ice; the more pure the ice, the larger $E_a$ and the smaller
$\kappa$. Because these impurities are hard to quantify and control, 
the static conductivity of water ice varies by orders of magnitude ($10^{-7}$ to $10^{-10}$ S m$^{-1}$ at $T=263$ K
for single crystals) between different studies \citep{pet93}. Likewise, 
\citet{sti13} report $\kappa = 10^{-9}$ to $10^{-10}$ S m$^{-1}$ for meteoric ice cores from Antarctica
and Greenland for temperatures between 210 and 220 K.

\citet{shi12} created amorphous water ice films of 1000 to 3000 monolayers (corresponding to 300 to 900 nm thickness)
on a microbalance and verified that the ice did not charge up
when the ion penetration depth reached the ice layer thickness, i.e., the energetic ions traversed the ice and hit the
gold substrate of the microbalance underneath the ice. For ice films thick enough to charge
up, the authors found that the discharging followed an exponential decrease with two
different decay constants, representing deep and shallow charges in the ice.
The half-life of the shallow charges ($\sim400$ s) was independent
of ice temperature, but the much longer half-life of the deep charges decreased with temperature.
\citet{bu15} studied the surface potential and proton transport in similar water ice films 
and found a constant permittivity for the ice between 30 and 95 K.
In some cases \citet{shi12} also observed a dielectric breakdown 
once the surface voltage exceeded hundreds of Volt. We do not expect this phenomenon for our experiments
because our ice sample is 10,000 times thicker. Such breakdowns may, however, contribute to the strong signals of frost 
being detached from metal surfaces (see Section \ref{sec:frost}).

\subsection{Ion sputtering yield}

The experiment set-up (Fig.~\ref{fig:schematic}) allows us to assess the sputtering yield from the 
ion beam hitting the ice surface by monitoring the total pressure in the vacuum chamber.
If we assume that a constant particle source $q_s$ in addition to the residual
outgassing rate $q_r$ is activated at time $t_0$
and turned off again at time $t_1$ we expect the following evolution of the chamber pressure $p(t)$:
\begin{equation}
p(t) = \frac{q_{r}}{S} + \frac{q_{s}}{S} \left(1-\exp\left(-\frac{S}{V}t\right) \right), \textup{ for } t_0<t<t_1
\label{eq:pressureequation1}
\end{equation}
\begin{equation}
p(t) = \frac{q_{s}}{S} \exp\left(-\frac{S}{V}t\right) + \frac{q_r}{S}, \textup{ for } t\geq t_1
\label{eq:pressureequation2}
\end{equation}
with $V = 1.6$ m$^3$ the volume of the chamber and antechamber in Fig.~\ref{fig:schematic},
$S = 0.35\pm0.1$ $m^{3}$ s$^{-1}$ the effective pumping speed of the turbomolecular pump, and the temperature 
of the chamber walls $T = 300$ K. 
Ejected particles that happen to re-impact a cold surface (the cooling plate or the nitrogen-conducting tubes) will be cold trapped
and thus not appear in the observed pressure rise. We neglected this effect since the ratio of cooled surfaces
to surfaces at 300 K amounts to less than 2\%.
The time scale established from Eq.~\ref{eq:pressureequation2} is $\tau_{eq} = 1/\lambda \approx 5$ s.
The new equilibrium pressure reached due to sputtering, $p(t_1) = p(t_0) +\Delta p = q_{r}/S + q_{s}/S$,
is linked to the sputtering yield $Y$ (ratio of ejected molecules per impacting ions) by the ideal gas law:
\begin{equation}
Y = \frac{\Delta p S}{kT} \frac{e^{-}}{i} 
\label{eq:eq}
\end{equation}
The number of impacting ions is easily measured via the ion beam current $i$ and $e^{-}$ denotes the elementary charge.

To relate our sputtering yields to previous experimental studies we will rely on the semi-empirical
formula derived by \citet{fam08}. It serves as a summary of previous sputtering experiments 
with compact water ice films.
For ion energies below 1 keV, the sputtering yield of ions in water ice 
can be described by a cascade of elastic collisions, whereas 
at higher ion energies the so-called electronic sputtering dominates.
The total sputtering yield (water molecules per incident ion) is the sum 
of the two contributions. \citet{fam08}
derived an expression including both contributions, 
which fit their laboratory measurements and results of other research groups \citep{joh10}
for H$^{+}$, He$^{+}$, N$^{+}$, O$^{+}$, Ne$^{+}$, and 
Ar$^{+}$ beams:
\begin{equation}
Y(E,m_1,Z_1,\theta,T) = \frac{1}{U_0}\left( \frac{3}{4\pi^2C_0}\alpha S_n+\eta S^{2}_{e}\right)
    \left(1+\frac{Y_1}{Y_0}\exp\left(-\frac{E_a}{kT}\right)\right) \cos^{-f}(\theta)
\label{eq:sputteryield_fama}
\end{equation}
Equation \ref{eq:sputteryield_fama} quantifies the sputtering yield as a sum of elastic and electronic
sputtering, described by the nuclear stopping power $S_n(E,m_1,Z_1)$ and the electronic stopping power $S_e(E,m_1,Z_1)$. 
The sputtering yield depends on energy $E$, mass of impactor $m_1$, 
atomic number of impactor $Z_1$, the incidence angle $\theta$ from the surface normal 
($60^{\circ}$ in our experiment set-up), and temperature $T$. 
For the sublimation energy of water, $U_0$, \citet{fam08} assumed 0.45 eV.
The effective cross-section for low energy recoils, $C_0=1.3$ \AA$^{2}$,
the activation energy, $E_a = 0.06$ eV, and
$Y_1/Y_0=220$ are constants. The parameter describing the
angular dependence calculates to $f=1.78$ for Ar$^{+}$.
From the angular dependence in Eq.~\ref{eq:sputteryield_fama} one expects an order of magnitude 
higher sputtering yields at ion incidence angles around 80$^{\circ}$ than for 
perpendicular ion impacts. The condition is that the ice sample is microscopically smooth. \citet{kue98} studied graphite
surfaces of varying roughness on a $\mu$m scale and found that the sputtering yield increased only by a factor of
2.5 when the ion incidence angle increased from 0$^{\circ}$ to 80$^{\circ}$. 
For a smooth graphite surface, they confirmed that $Y$ increases by more than a decade. 
In the following section, we will compare our new experiment results for Ar$^{+}$ to the predictions 
in Eq.~\ref{eq:sputteryield_fama}. 

\section{Results and discussion}\label{sec:results}

We first describe the measured charging and discharging time scales for the ice
before proceeding to our first results on the ion and electron sputtering.

\subsection{Surface charging}\label{sec:charging}

\begin{figure}
\begin{center}
\includegraphics[width=1.0\textwidth]{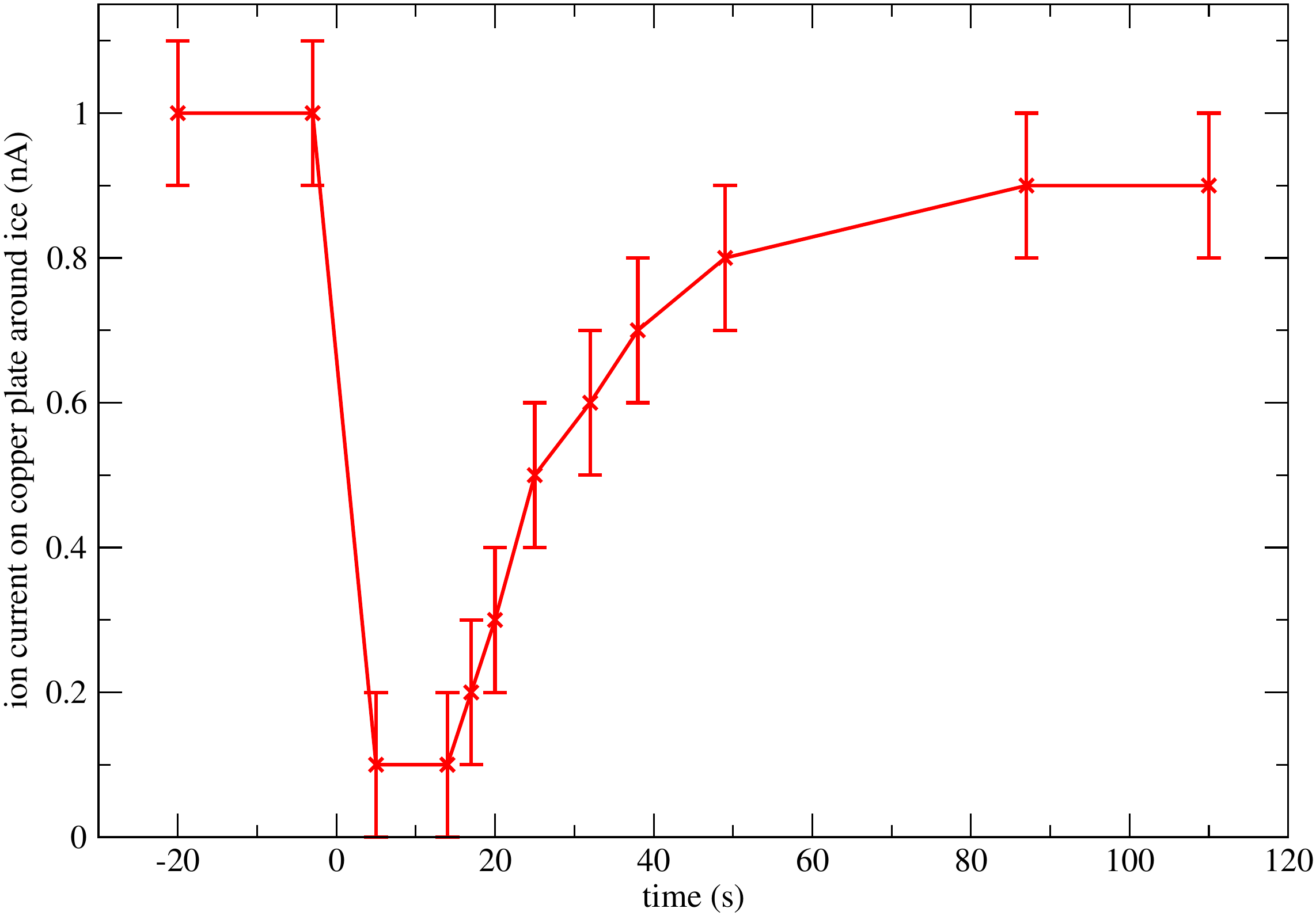}
\end{center}
\caption{Charging of the ice surface during ion bombardment for the case of 10 keV Ar$^{+}$ ions impacting ice of $T=142$~K. 
At the time $t=0$ s, the ion beam
is released at the centre of the ice sample. The ion current measured on
the copper ring surrounding the ice sample then increases from zero to almost the full current with a half-life
for charging of $t_{c,1/2} = 21\pm4$ s.}\label{fig:charging}
\end{figure}

When the ion beam was directed at the centre of the ice sample, the ion current measured
on the copper ring initially dropped to zero. After a few seconds, depending on the temperature
of the ice, the ion current increased to reach a stable value
of 80\% to 90\% of the ion current measured when the beam was directed at the copper ring.
This indicated that most ions were deflected by the surface potential off their original 
line of flight onto the copper ring.
Figure~\ref{fig:charging} shows a typical example of this surface charging
for a 10 keV Ar$^{+}$ beam with a total current of $i=1.1\pm0.1$ nA directed at ice of $T=142$~K.
The measured charge versus time can be described by an exponential increase
\begin{equation}
i(t) = i_0 (1-\exp(-\lambda_c t))
\label{eq:charging}
\end{equation}
with the charging time constant $\lambda_c$. 
Table \ref{tab:charging} lists the half-life times $t_{c,1/2}$ = $\ln(2)/\lambda_c$ 
for ions to charge the ice sample for the five combinations
of ice temperatures and ion energies studied.
The half-life was defined by the time when half of the ion beam current
was measured on the copper ring surrounding the ice. The ice sample with higher porosity
and temperature took significantly longer to charge up.

\begin{table}
\begin{center}
\caption{Measured properties of thick porous ice sample. The columns denote energy of Ar$^{+}$ ions,
ion current, ice temperature, ice density, half-life for surface charging, and ion sputtering yield (water molecules
per incident ion).}\label{tab:charging}
\begin{tabular}{rlrlrl}
\hline
$E$ (keV) & $i$ (nA) & $T$ (K) & $\rho$ (g cm$^{-3}$) & $t_{c,1/2}$ (s) & $Y$\\
3 &$1.25\pm0.15$ &$143$ &0.2 & $14\pm2$ & $< 100$\\
10&$1.1\pm0.1$ &$142$& 0.2 & $21\pm4$ & $250 \pm 125$\\
3 &$1.2\pm0.2$ & $116$& 0.29 & $5\pm1$ & $< 110$\\
10 &$1.6\pm0.2$ &$116$& 0.29 & $8\pm1$ & $< 85$\\
30& $1.6\pm0.2$ &$116$& 0.29 & $6\pm1$ & $60\scriptsize{\begin{array}{rr}+85 \\-60\end{array}}$\\
\end{tabular}
\end{center}
\end{table}

The fraction of the ion beam not hitting the ice follows Equation \ref{eq:fullycharged}
for a plane parallel capacitor charging up. However,
the linear increase of charging time with ion energy expected from
Equation \ref{eq:effective_potential} was not observed in measurements.
The asymptotic rise to an equilibrium also agrees with the observations by \citet{shi10,shi12} for compact water ice films.
The observed charging timescales on the order of 10 seconds are similar
to our simple model of a capacitor, in contrast to the 100--1000 s observed for a water ice film \citep{shi10}. 
The results in Table \ref{tab:charging} are ambiguous because the change in timescale could be attributed
to the density or the temperature change. 
But since the permittivity of dry snow increases with density \citep{mae06}, we would
expect that more compact ice should take longer to charge up (Equation \ref{eq:fullycharged}). 
We therefore deem it more likely that the temperature difference between the two ice samples
caused the different charging timescale. Such a temperature effect on electric properties of the ice 
does not follow from observations of compact water ice \citep{mae06}. Since we cannot quantitatively predict
this effect, the target area of the ion beam had to be monitored simultaneously 
during all ion sputtering experiments. The signal caused by ion sputtering (see Section \ref{sec:yield}) had to be determined 
in the first seconds ($t< t_{c,1/2}$) during ion bombardment.

\subsection{Discharging of the ice sample}\label{sec:discharging}

The ice surface must be discharged if we want to perform unbiased
ion sputtering experiments. The electron gun achieved this task, 60 seconds at an electron energy of 50 or 100 eV 
proved sufficient to completely discharge the surface. Subjecting the ice surface to a longer
time of electron irradiation did not have an impact on the behaviour of measured
ion beam current versus time in subsequent ion sputtering experiments.
We also studied the natural discharge time of the ice sample without electron irradiation.
The discharging time increased at low temperatures, as found by \citet{shi12}
for compact water ice films. The decay times observed for thick porous ice were the same order of magnitude
as the short decay component found by \citet{shi12}. 
Our measurements did not allow the distinction of two separate decay constants as described by \citet{shi12}.

\begin{figure}
\begin{center}
\includegraphics[width=1.0\textwidth]{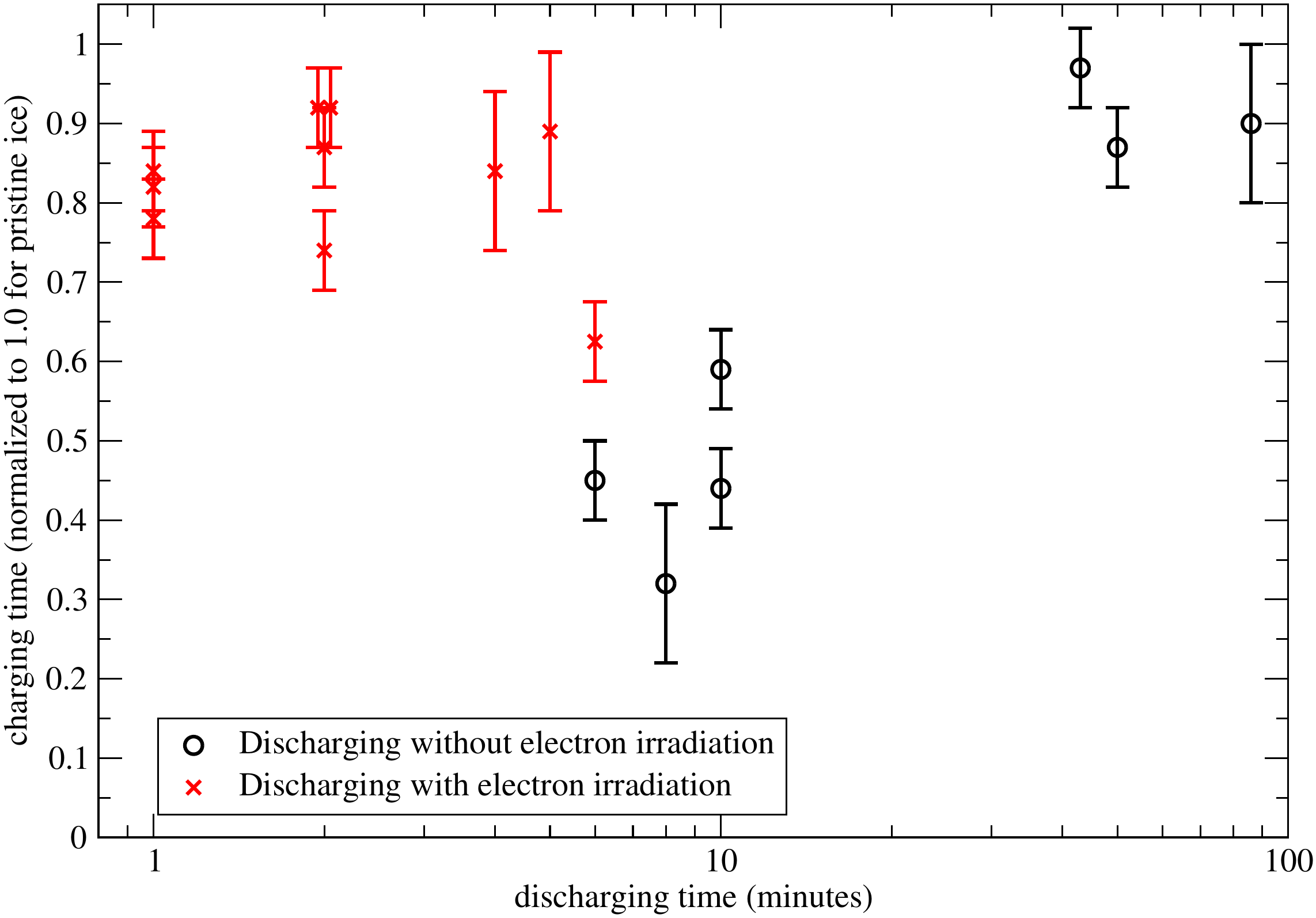}
\end{center}
\caption{The effect of electron irradiation on surface charging. 
Red ``x'' symbols: experiments after electron gun had been
running to neutralize ice surface; black circles: electron gun inactive.}\label{fig:discharging}
\end{figure}

Since we could not measure the ice potential directly, we had 
to resort to an indirect measurement. After fully 
charging the ice surface we waited for 1 to 100 minutes before directing the ion 
beam again at the ice. If the surface had been completely discharged in the meanwhile, 
the new charging time was identical to the one observed for pristine ice. 
Otherwise the new charging time was only a fraction (0.0 to 1.0) of the original charging time.
Figure~\ref{fig:discharging} illustrates the natural discharge times without
simultaneous electron irradiation (black circles)
and the effect of electron irradiation on discharging (red ``x'' symbols). 
In both cases, charging times are plotted against the discharging
times in minutes. 
The charging times are normalized to 1.0 for pristine ice.  
The results of the discharging experiments without electron irradiation are summarized in Table \ref{tab:discharging}. 
The experiments at cold temperatures (89 K) were performed without the copper ring
around the ice sample, therefore the surface charging properties
could not be measured accurately. This value for the discharging time is only an upper limit.

\begin{table}
\begin{center}
\caption{Half-life $t_{d,1/2}$ of the ice surface discharging without electron irradiation. Note that the
times are given in minutes.}\label{tab:discharging}
\begin{tabular}{rrr}
\hline
$T$ (K) & $\rho$ (g cm$^{-3}$) & $t_{d,1/2}$ (minutes)\\
$142$ & 0.2 & $4.6\pm0.9$\\
$116$ & 0.29 & $23\pm8$\\
$89$ & 0.29 & $>60$\\
\end{tabular}
\end{center}
\end{table}

Our results imply via Equation \ref{eq:kappa} that the static conductivity of porous water ice
decreases from $\kappa = 3\times10^{-14}$ S m$^{-1}$
at 142 K to $7\times10^{-15}$ S m$^{-1}$ at 116 K. By comparison,
the conductivity of pure deionized water at room temperature is on the order of $10^{-5}$ S m$^{-1}$
and the mean conductivity of the oceans on Earth is 3.27 S m$^{-1}$ \citep{kay05}. If we assume
that the conductivity of the porous ice depends on temperature according to $\sim \exp(-E_a/k_BT)$ \citep{sti13,pet93}, this
implies an activation energy $E_a$ of 0.08 eV.
At temperatures representative for the Galilean moons, the static conductivity of a frost-covered
porous ice surface is on the order of $10^{-15}$ S m$^{-1}$ or even lower.
This means that an icy moon without ionosphere can be considered a non-conducting obstacle to the surrounding plasma. 
If such a moon has neither an intrinsic magnetic field nor slows down the plasma by mass loading
\citep{jia15}, it would create a lunar-type of wake in the plasma \citep{hal15}. 
Ganymede, however, has a magnetic field \citep{kiv97} and Europa, Ganymede, and Callisto also
have ionospheres with $n_e = 10^9$ to $10^{10}$ m$^{-3}$ 
peak electron density close to the surface \citep{kli97,evi01,kli02}. 
The surface charging observed in laboratory experiments
therefore will not take place on a macroscopic scale on these moons. Any spot exposed to ion irradiation will 
also be exposed to electrons and solar UV light. The actual surface potential
will be determined by the balance of incoming and outgoing currents.

\subsection{Electron sputtering}

The signal caused by electrons sputtering the water frost
on the metal surface was well visible via a pressure rise in the chamber. 
The height of the pressure rise depended on the energy and the flux of the electrons.
The signal implied $Y=0.3\pm0.15$ water molecules per incident electrons at 100~eV via Equation \ref{eq:eq}. 
The pressure rise observed with the cold cathode gauge at a lower precision (see Section \ref{sec:procedure}) 
yielded $Y=0.3\pm0.3$. This independent measurement 
at another place in the vacuum chamber confirmed the validity of the pressure rise
approach. However, we did not yet quantify the electron sputtering yield as a function
of electron energy. We also will have to narrow the electron beam
to compare the sputtering signal from
the ice sample against the signal from the frost on metal surfaces. This comparison
will decide if the electron sputter behaviour depends on ice properties.
In the present study, we used a completely defocused electron beam to ensure surface neutralization. 
Contrary to ion irradiation, the sputtering caused by electrons does not choke itself off because
of the secondary electrons emitted from the surface. For electron energies between roughly 50 eV and 10 keV
the sputtering yield of secondary electrons is larger than 1.0 \citep{jur95}, thus more electrons will leave the ice
than impacting the ice and newly arriving electrons will not be deflected
off the surface. The (positive) surface potential due to electron irradiation will not exceed a few Volt because
secondary electrons have energies around a few eV almost independent of the energy of the impactors \citep{jur95}. 
This remaining potential is irrelevant for a subsequent sputtering experiment with keV-energy ions. They
cause a positive surface potential three orders of magnitude
higher (see Section \ref{sec:theory_charging}) than the few Volt due to secondary electrons.

\subsection{Ion sputtering}\label{sec:yield}

The surface charging must be taken into account for an unbiased ion sputtering
experiment. When the ice surface is charged the majority of the ions will not hit the ice 
sample but rather the metal surfaces and frost around the sample. 
To measure ion sputtering we directed the ion beam at the ice until 
the surface potential had reached an equilibrium, 
discharged the ice during one minute with electrons and repeated the measurement. 
This way we avoided waiting times of one hour after each experiment (see Table \ref{tab:discharging}).   

For the cold ice ($T=116$ K), the charging timescale was too short and the sputtering yield
was too low to observe a rise of the chamber pressure significantly higher than the precision
of the pressure gauge for all ion energies. 
From Equation \ref{eq:eq} it follows that $Y<150$ molecules per ion (see last column in Table \ref{tab:charging}).

\begin{figure}
\begin{center}
\includegraphics[width=1.0\textwidth]{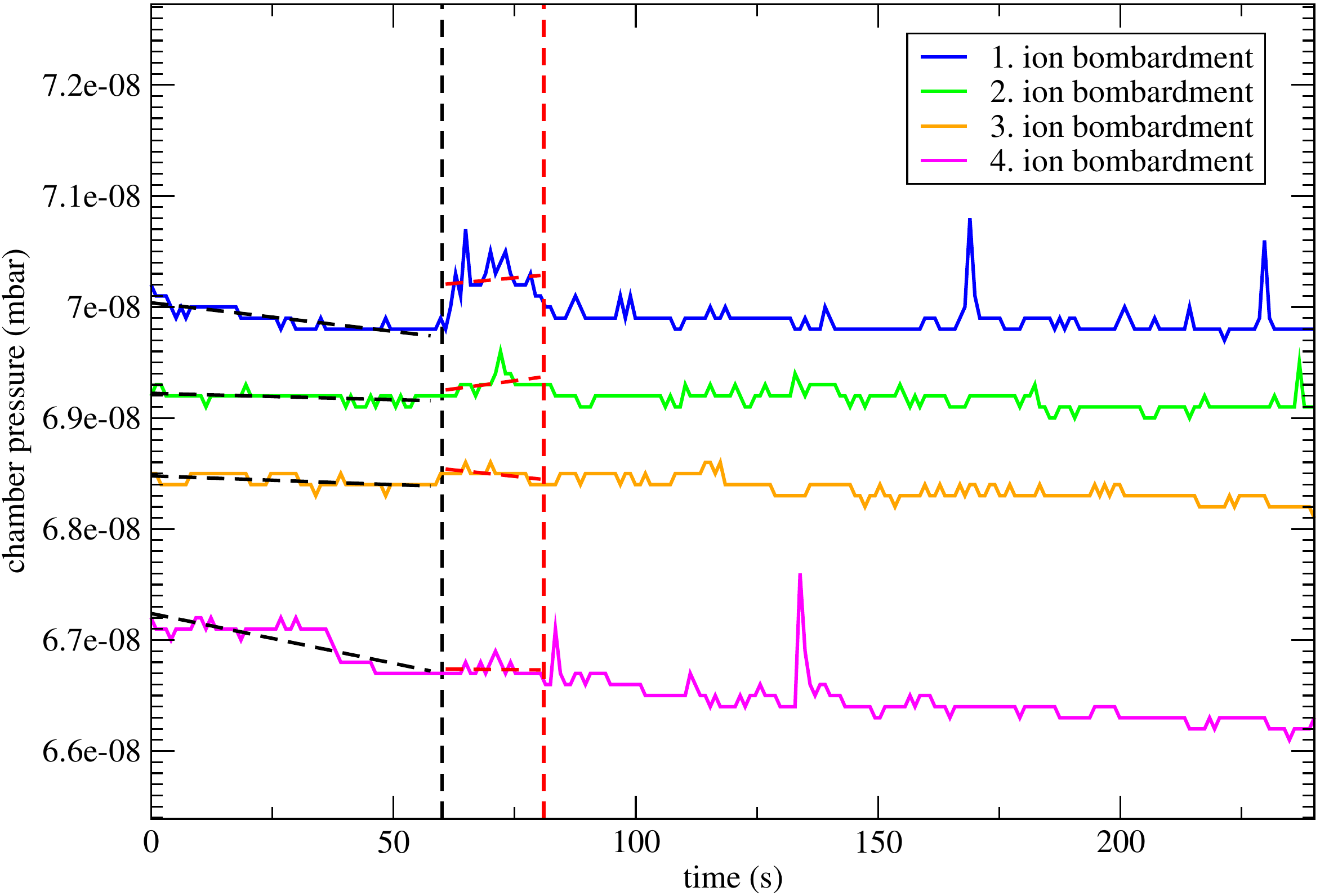}
\end{center}
\caption{Pressure rise observed due to ion sputtering. Shown are four of nine observations
for a 10 keV Ar$^{+}$ ion beam directed at the centre of the ice sample at $T=142$ K.
The black vertical line denotes the onset of sputtering, the red vertical line denotes
the moment when one half-life of surface charging has passed. The pressure spikes
at 130 and 170 s are due to frost being sputtered off the metal surfaces around the sample.}\label{fig:p(t)_warmice_10_keV}
\end{figure}

For warmer temperatures ($T=142$ K) and 10 keV ion energy,
we observed a pressure rise in nine consecutive observations. It occurred within the first 3 seconds
after the onset of ion sputtering. 
Figure~\ref{fig:p(t)_warmice_10_keV} shows four of these nine measurements as chamber pressure
versus time. The black vertical line indicates the moment when the ion beam was directed
at the centre of the ice, the red vertical line indicates the time when one charging half-life (21 s)
has passed. After that time mark, irregular pressure spikes could occur as long as the ion beam was active.
We explain this effect by the ion beam being deflected off the sample and hitting frost deposited on metal surfaces
(see Section~\ref{sec:frost} for more details). 
We could not verify the shape of the pressure rise predicted in Equation \ref{eq:pressureequation1}
because the precision of the pressure gauge of $10^{-10}$ mbar was close to the pressure rise $\Delta p$. 
But linear regressions to the pressure versus time in the minute before ion sputtering (black dashed
lines) and for the first 21 seconds during ion sputtering (red dashed lines) 
allowed us to estimate the $\Delta p$ needed in Equation \ref{eq:eq}. 

\begin{figure}
\begin{center}
\includegraphics[width=1.0\textwidth]{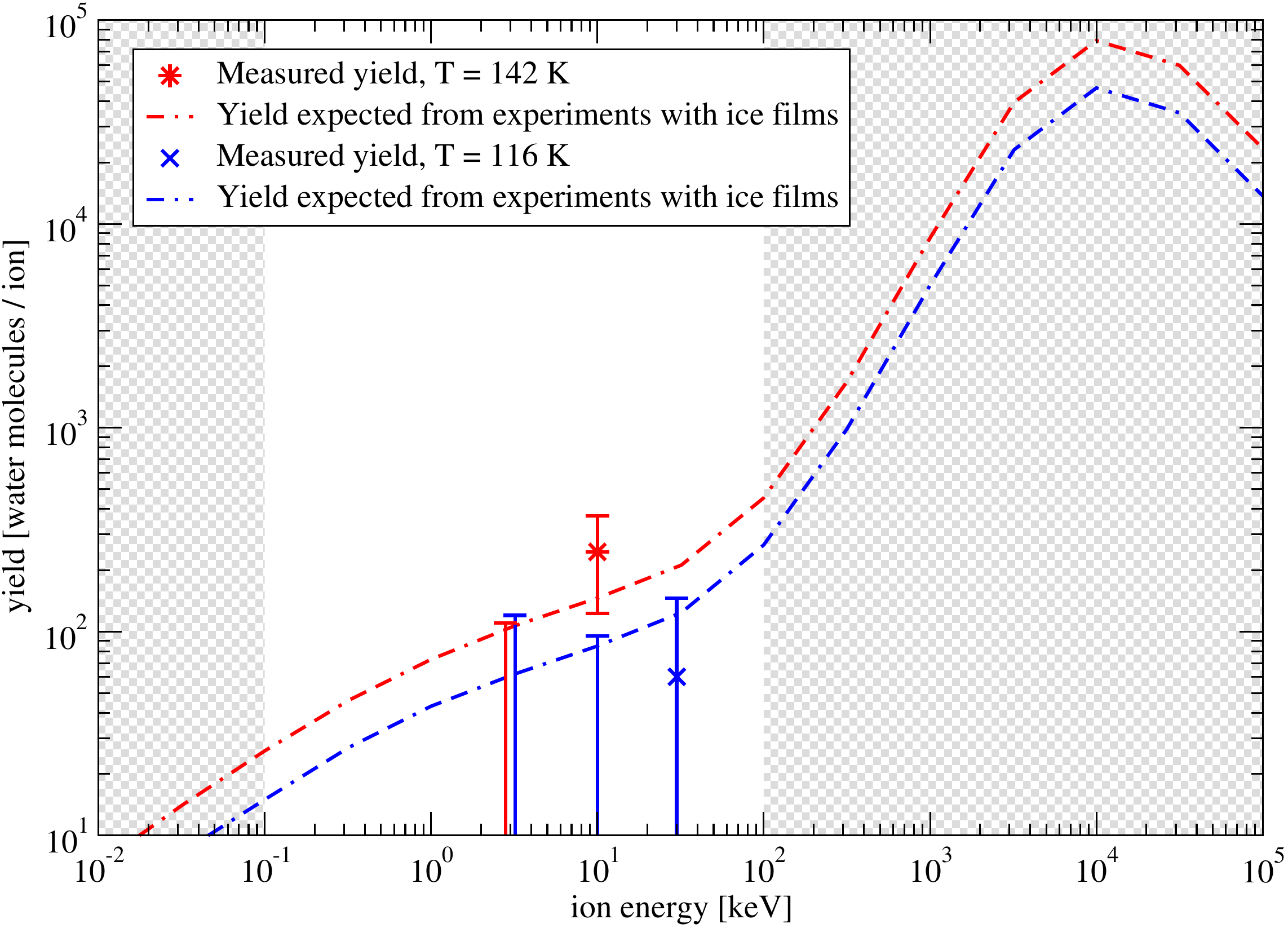}
\end{center}
\caption{Sputtering yield for Ar$^{+}$ ions from porous water ice. The results
obtained with our experiment set-up are compared with the semi-empirical formula based on previous laboratory experiments
with compact water ice films provided by \citet{fam08}.}\label{fig:fama}
\end{figure}

From Fig.~\ref{fig:p(t)_warmice_10_keV} we obtain an average $\Delta p=(2\pm1)\times10^{-10}$ mbar
for all nine measurements. This implies
a sputtering yield of $Y=250\pm125$ according to Equation \ref{eq:eq}. The probability
that we mistook random fluctuations in chamber pressure for a signal nine times in a row is 
$0.5^9 < 0.2\%$. For the lower energy of 3 keV, we only could derive an upper limit of $Y<100$.  
In Fig.~\ref{fig:fama} we plot
the observed sputtering yields or their upper limits against the sputtering yields expected
from previous ice sputtering experiments performed with water ice films on a microbalance 
(Equation \ref{eq:sputteryield_fama}).
Our results are consistent with the notion that sputtering from porous regolith
ice is similar to sputtering from compact monolayers of amorphous water ice.

The basic limitation of the current set-up is that only intense sputtering signals with a yield
larger than 100 can be detected due to the limits imposed by the residual pressure in the chamber
and the precision of the pressure measurement. This means we can only identify sputtering yields
for ion energies larger than 10 keV at ice temperatures considered here. Higher ice temperatures would 
result in a higher sputtering yield, but to simulate the icy moons of Jupiter, temperatures between
80 K and 150 K are mandatory (see Section \ref{sec:theory}).
In future experiments we will combine our set-up with a cooled microbalance opposite the ice sample to
collect sputtered material and thus to enlarge our measurement range.
Addition of a microbalance will also allow us to directly compare our
results to previous ice sputtering experiments with thin water ice layers on a microbalance
\citep{shi95,orl03,joh04,fam08,cas10,shi12}.

\subsection{Sputtering of irregular frost on metal surfaces}\label{sec:frost}

\begin{figure}
\begin{center}
\includegraphics[width=1.0\textwidth]{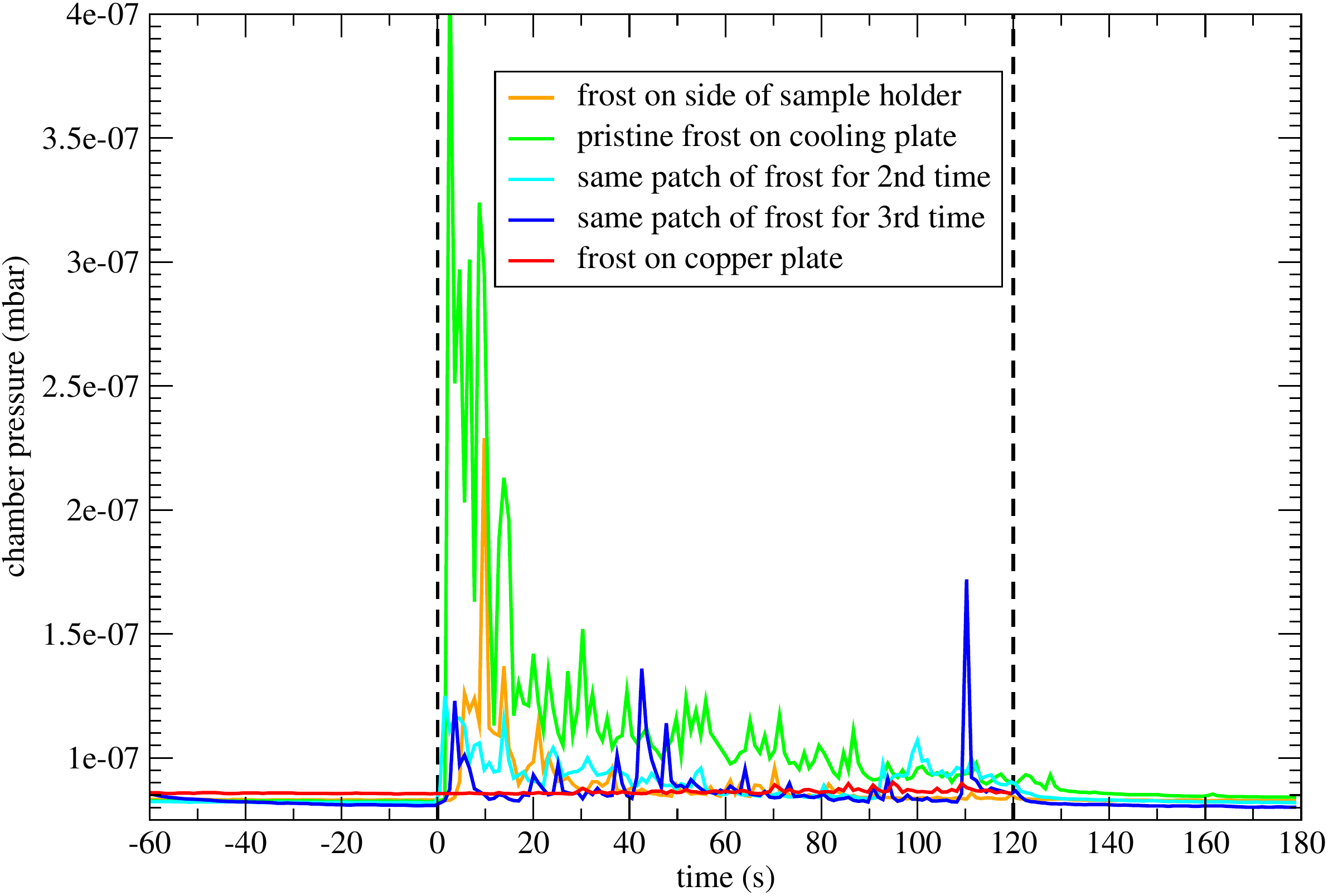}
\end{center}
\caption{Pressure rise observed due to 10 keV Ar$^{+}$ ions hitting thin irregular frost on various metal
surfaces. Orange: frost on the outer bottom of the sample holder ($T=90$ K),
green, cyan, and blue: frost on the cooling plate ($T=101$ K),
red: thin frost on the copper ring above the ice sample ($T>142$ K).
The first black vertical line denotes the onset of sputtering, the second line denotes
the moment when the ion beam is blocked.}\label{fig:p(t)_frost}
\end{figure}

We emphasize the importance of studying and monitoring surface charging of ice during any
ion sputtering experiment. Figure~\ref{fig:p(t)_warmice_10_keV} demonstrated the danger
of mis-interpreting strong signals of detached frost as the regular ice sputtering signal. This
became more obvious when we directed the ion beam from the beginning at the copper ring or other frost covered 
metal surfaces. Figure~\ref{fig:p(t)_frost} shows the chamber pressure
for five of those cases, the ion beam starts hitting the frost covered metal surfaces at the time $t = 0$.
The pressure spikes appear within the first few seconds afterwards and may be up to 1000 times 
stronger than the nominal sputtering signal if the frost on the cooling plate is targeted. 
The frost signal from the copper ring is much weaker (red line in Fig.~\ref{fig:p(t)_frost}) 
but still stronger than the sputtering signal from the ice sample.
The pressure fluctuates rapidly, but the spikes tend to become weaker during ion bombardment and 
if the same experiment is repeated for the same patch of frost on metal. Our interpretation is that entire flakes of 
frost are detached from the metal surfaces by electrostatic repulsion. 
The quadrupole mass spectrometer shows that the pressure
rise during such incidences is solely due to H$_{2}$O molecules.
The process sets in too fast and shows too large variations
to be attributed to the ion beam warming up the frost. 
This type of intense frost sputtering does not apply
to the surfaces of icy moons because there is no conducting metal surface directly underneath the top frost layer.  
Given these observations, we consider a thick ice layer crucial for sputtering experiments 
to electrically decouple the ion charges from the metal surfaces underneath. 

\section{Conclusions}\label{sec:conclusion}

We have presented first experimental studies of ion sputtering in thick and porous water ice representative for the surface
of icy moons. We have shown that the surface charging of the ice must be monitored during the experiment
to correctly interpret the observed sputtering yield. The apparent signal obtained when a thin or irregular
frost layer is detached from a metal surface due to electrostatic repulsion may be orders of magnitude
larger than the actual sputtering rate of water ice. Electric conductivity of porous water ice regolith
at temperatures close to 100 K is very low, on the order of $10^{-15}$ S m$^{-1}$, even if NaCl is present. 

Since a thick and porous ice layer cannot be attached to the tip of a microbalance we resorted
in these first experiments to monitoring the rise and fall of the chamber pressure during
sputtering experiments. This pressure rise method leads to sputtering yields similar to previous 
laboratory experiments performed with thin water ice films sputtered off a microbalance.
Our results are thus consistent with the notion that sputtering from porous regolith
ice is similar to compact monolayers of water ice. But more experiments at various energies, incidence
angles, and different ion species are required to make a stronger statement.
The residual pressure in the chamber and the measurement precision prevent us from measuring
sputter yields less than 100. As a result, we only obtain upper limits for the sputtering yield
at energies below 10 keV.

Our experiments have, with the limitations stated above, three main implications 
for surface and exosphere models of the Galilean moons. So far, the sputtering yields
implemented in exosphere models need not be changed. However, plasma models must take into
account that the electric conductivity of the surface is very low. As a consequence, electron irradiation
may have to be included explicitly in exosphere models.

\section*{Acknowledgements}

The work by A. Galli and A. Pommerol at the University of Bern
has been supported by the Swiss National Science Foundation. The
work in this paper has been partially performed in the context of
the activities of the ISSI International Team Nr. 322, www.issibern.
ch/teams/exospherejuice/. We also would like to thank G. Bodmer,
J. Gonseth, and A. Etter for their relentless support of the scientific
work at the MEFISTO facility and two anonymous referees whose
comments helped improve the paper.


\section*{References}

\begin{thebibliography}{}

\bibitem[Andreas(2007)]{and07}Andreas, E.L., 2007. New estimates for the sublimation rate for ice on the Moon. 
Icarus 186, 24--30.

\bibitem[Barth et al.(1997)]{bar97}Barth, C.A., Hord, C.W., Stewart, A.I.F., Pryor, W.R., Simmons, 
K.E., McClintock, W.E., Ajello, J.M., Naviaux, K.L., and Aiello, J.J., 1997. 
Galileo ultraviolet spectrometer observations of atomic hydrogen in the atmosphere of Ganymede.
Geophysical Research Letters, 24, 2147. 

\bibitem[Brocard(2008)]{bro08} Brocard, E. 2008. Ground-based        remote        sensing        of        the        troposphere        in
the        thermal        infrared. PhD thesis, Philosophisch-Naturwissenschaftliche Fakult\"at, University of Bern, Bern, Switzerland.

\bibitem[Brouet et al.(2016)]{bro16} Brouet, Y. et al. 2016. Measurements of permittivity of porous water ice. 
\textit{Submitted to Journal of Geophysical Research}.

\bibitem[Brown and Hill(1996)]{bro96}Brown, M.E., Hill, R.E., 1996. Discovery of an extended sodium atmosphere around Europa. 
Nature 380, 229.

\bibitem[Brown et al.(2001)]{bro01}Brown, M.E., 2001. Potassium in Europa’s atmosphere. 
Icarus 151, 190.

\bibitem[Bu and Baragiola(2015)]{bu15}Bu, C., Baragiola, R.A. 
Proton transport in ice at 30--140 K: Effects of porosity. The Journal of Chemical Physics 143, 074702.

\bibitem[Calvin et al.(1995)]{cal95}Calvin, W.M., Clark, R.N., Brown, R.H.,
Spencer, J.R., 1995. Spectra of the icy Galilean satellites from 0.2 to 5 $\mu$m:
A compilation, new observations, and a recent summary. Journal of Geophysical Research, 100, 19041--19048.

\bibitem[Carlson(1999)]{car99}Carlson, R.W., 1999. A Tenuous Carbon Dioxide Atmosphere 
on Jupiter's Moon Callisto. Science, 283, 820.

\bibitem[Cassidy and Johnson(2005)]{cas05}Cassidy, T.A., Johnson R.E., 2005. 
Monte Carlo model of sputtering and other ejection processes within a regolith. Icarus 176, 499--507.

\bibitem[Cassidy et al.(2010)]{cas10}Cassidy, T., Coll, P., Raulin, F., Carlson, R.W., Johnson, R.E., Loeffler, 
M.J., Hand, K.P., Baragiola, R.A., 2010. Radiolysis and Photolysis of Icy Satellite Surfaces: 
Experiments and Theory. Space Science Reviews. doi:10.1007/s11214-009-9625-3.

\bibitem[Cassidy et al.(2013)]{cas13}Cassidy, T.A., Paranicas, C.P., Shirley, 
J.H., DaltonIII, J.B., Teolis, B.D., Johnson, R.E., Kamp, L., Hendrix, A.R., 2013.
Magnetospheric ion sputtering and water ice grain size at Europa. Planetary and Space Science 77, 64--73. 

\bibitem[Domingue and Verbiscer(1997)]{dom97}Domingue, D.L., Verbiscer, A., 1997. Re-analysis of the solar phase curves of the icy
Galilean satellites. Icarus 128, 49--74.

\bibitem[ESA(2014)]{ESA14}ESA, 2014. JUpiter ICy moons Explorer -- Exploring the emergence of habitable worlds
around gas giants -- Definition Study Report. ESA/SRE(2014)1, September 2014, 
http://sci.esa.int/juice/54994-juice-definition-study-report.

\bibitem[Eviatar et al.(2001)]{evi01}Eviatar, A., Vasyliunas, V.M., Gurnett, D.A., 2001. The ionosphere of Ganymede.
Planetary and Space Science 49, 327--336.

\bibitem[Fam\'{a} et al.(2008)]{fam08}Fam\'{a}, M., Shi, J., Baragiola, R.A., 2008. Sputtering of ice by low-energy ions. 
Surface Science 602, 156.

\bibitem[Galli et al.(2015)]{gal15}Galli, A., Pommerol, A., Wurz, P., Jost, B., Scheer, J.A., Vorburger, A., Tulej, M.,
Thomas, N., Wieser, M., Barabash, S., 2015, Realistic ice sputtering experiments for the surfaces of
Galilean moons, EPSC Abstracts, 10, EPSC2015-771-1.

\bibitem[Grundy et al.(1999)]{gru99}Grundy, W.M., Buie, M.W., Stansberry, J.A., Spencer, J.R., 1999. 
Near-Infrared Spectra of Icy Outer Solar System Surfaces: Remote Determination of H$_{2}$O Ice Temperatures. Icarus 142, 536--549.

\bibitem[Halekas et al.(2015)]{hal15}Halekas, J.S., Brain, D.A., Holmstr\"om, M. 2015. Moon's Plasma Wake. In: 
Keiling, A., Jackman, C.A., Delamere P.A. (Eds.), Magnetotails in the Solar System. Geophysical Monograph 207,
American Geophysical Union and John Wiley \& Sons, Inc., Washington D.C., USA. 

\bibitem[Hall et al.(1995)]{hal95}Hall, D.T., Strobel, D.F, Feldman, P.D., McGrath, M.A., Weaver, H.A., 1995. 
Detection of an oxygen atmosphere on Jupiter's moon Europa. Nature 373, 677.

\bibitem[Hall et al.(1998)]{hal98}Hall, D.T., Feldman, P.D., McGrath, M.A., Strobel, D.F., 1998. 
The Far-Ultraviolet Oxygen Airglow of Europa and Ganymede. The Astrophysical Journal 499, 475.

\bibitem[Hansen et al.(2005)]{han05}Hansen, C.J., Shemansky, D.E., Hendrix, A.R., 2005. 
Cassini UVIS observations of Europa’s oxygen atmosphere and torus. Icarus 176, 305.

\bibitem[Hohl(2002)]{hoh02}Hohl, M., 2002. MEFISTO II: design, setup, characterization and
operation of an improved calibration facility for solar
plasma instrumentation. PhD Thesis, University of Bern.

\bibitem[Hohl et al.(2005)]{hoh05}Hohl, M., Wurz, P., Bochsler, P., 2005.
Investigation of the density and temperature of electrons in a compact
2.45 GHz electron cyclotron resonance ion source plasma by x-ray measurements. 
Plasma Sources Science and Technology 14, 692--699. doi:10.1088/0963-0252/14/4/008.

\bibitem[Jia(2015)]{jia15}Jia, X. 2015. Satellite Magnetotails. In: 
Keiling, A., Jackman, C.A., Delamere P.A. (Eds.), Magnetotails in the Solar System. Geophysical Monograph 207,
American Geophysical Union and John Wiley \& Sons, Inc., Washington D.C., USA. 

\bibitem[Johnson(1989)]{joh89}Johnson, R.E., 1989.
Application of Laboratory Data to the Sputtering of a Planetary Regolith. Icarus 78, 206--210.

\bibitem[Johnson et al.(2004)]{joh04}Johnson, R.E., Carlson, R.W., Cooper, J.F., Paranicas, C., Moore, M.H., Wong, M.C., 2004. 
Radiation effects on the surfaces of the Galilean satellites. In: Bagenal, F. (Ed.), 
Jupiter: Atmosphere, Satellites and Magnetosphere. University of Arizona Press, Tucson, USA.

\bibitem[Johnson and Liu(2010)]{joh10}Johnson, R.E. and Liu, M., 2010.
Sputtering of Surfaces, Sputtering Data for H$_{2}$O ice, http://people.virginia.edu/~rej/h2o.html 

\bibitem[Jost et al.(2016)]{jos16} Jost, B., Pommerol, A., Poch, O., Gundlach, B.,
Leboeuf, M., Dadras, M., Blum, J., Thomas, N., 2016. 
Experimental characterization of the opposition surge in fine-grained water-ice and high albedo analogs, 
Icarus 264, 109--131.

\bibitem[Jurac et al.(1995)]{jur95} Jurac, S., Baragiola, R.A., Johnson, R.E., and Sittler Jr., E.C., 1995.
Charging of ice grains by low-energy plasmas: Application to Saturn's $E$ ring.
Journal of Geophysical Research 100, A8, 14821--14831.

\bibitem[Kaye \& Laby(2005)]{kay05} Kaye \& Laby Online, 2005. Tables of Physical \& Chemical Constants (16th edition 1995). 
2.1.4 Hygrometry, Version 1.0, www.kayelaby.npl.co.uk.

\bibitem[Kivelson et al.(1997)]{kiv97} Kivelson, M.G., Khurana, K.K., Coronoti, F.V., Joy, S., Russell, C.T.,
Walker, R.J., Warnecke, J., Bennett, L., Polanskey, C., 1997. The magnetic field and magnetosphere of 
Ganymede. Geophysical Research Letters 24, 2155--2158.

\bibitem[Kliore et al.(1997)]{kli97} Kliore, A.J., Hinson, D.P., Flasar, F.M., Nagy, A.F.,
Cravens, T.E., 1997. The Ionosphere of Europa from Galileo Radio Occultations. Science 277, 355.

\bibitem[Kliore et al.(2002)]{kli02} Kliore, A.J., Anabtawi, A., Herrera, R.G., Asmar, S.W., Nagy, A.F.,
Hinson, D.P., Flasar, F.M., 2002. Ionosphere of Callisto from Galileo radio occultation observations.
Journal of Geophysical Research 107, 1407.

\bibitem[K\"ustner et al.(1998)]{kue98} K\"ustner, M., Eckstein, W., Dose, V., Roth, J, 1998.
The influence of surface roughness on the angular dependence of the sputter yield. Nucl. Instrum.
Methods B 145, 320--331.

\bibitem[Marconi(2007)]{mar07}Marconi, M.L., 2007. A kinetic model of Ganymede's atmosphere. Icarus 190, 155.

\bibitem[Marti et al.(2001)]{mar01}Marti, A., Schletti, R., Wurz, P., Bochsler, P., 2001.
Calibration facility for solar wind plasma instrumentation. Review of Scientific Instruments 72, 1354. doi:10.1063/1.1340020.

\bibitem[M\"atzler et al.(2006)]{mae06} M\"atzler, C., Ellison, W., Thomas, B., Sihvola, A., Schwank, M., 2006.
Dielectric properties of natural media. In: M\"atzler, C. (Ed.), Thermal Microwave Radiation: Applications for Remote Sensing''.
The Institution of Engineering and Technology, London, UK.

\bibitem[Noll et al.(1996)]{nol96}Noll, K.S., Johnson, R.E., Lane, A.L., Domingue, D.L., Weaver, H.A., 1996.
Detection of Ozone on Ganymede. Science, 273, 341.

\bibitem[Orlando and Sieger(2003)]{orl03}Orlando, T.M., Sieger, M.T., 2003. 
The role of electron-stimulated production of O$_{2}$ from water ice in the radiation processing of 
outer solar system surfaces. Surface Sciences 528, 1.

\bibitem[Petrenko(1993)]{pet93}Petrenko, V.F., 1993. Electrical Properties of Ice. 
Thayer School of Engineering, Dartmouth College, Hanover, NH, USA.

\bibitem[Plainaki et al.(2012)]{pla12}Plainaki, C., Milillo, A., Mura, A., Orsini, S., Massetti, S., Cassidy, T., 2012. 
The role of sputtering and radiolysis in the generation of Europa exosphere. Icarus 218, 956.

\bibitem[Rodriguez et al.(2009)]{rod09}Rodriguez, N.J., Rathbun, J.A., Spencer, J.R., 2009. 
Europa's thermal surface from Galileo PPR. $40^{\textup{th}}$ Lunar and Planetary Science Conference, Abstract, 2166. 

\bibitem[Rodriguez-Nieva et al.(2011)]{rod11}Rodriguez-Nieva, J.F., Bringa, E.M., Cassidy, T.A., 
Johnson, R.E., Caro, A., Fam\'{a}, M., Loeffler, M.J., Baragiola, R.A., Farkas, D., 2011. 
Sputtering from a porous material by penetrating ions. The Astrophysical Journal Letters 743, L5. doi:10.1088/2041-8205/743/1/L5.

\bibitem[Shi et al.(1995)]{shi95}Shi, M., Baragiola, R.A., Grosjean, D.E., Johnson, R.E., Jurac, S., Schou, J., 1995. 
Sputtering of water ice surfaces and the production of extended neutral atmospheres. Journal of Geophysical Research 100, 26387.

\bibitem[Shi et al.(2010)]{shi10}Shi, J., Fam\'{a}, M., Teolis, B.D., Baragiola, R.A., 2010. 
Ion-induced electrostatic charging of ice. Nucl. Instrum. Methods B 268, 2888--2891.

\bibitem[Shi et al.(2012)]{shi12}Shi, J., Fam\'{a}, M., Teolis, B.D., Baragiola, R.A., 2012. 
Ion-induced electrostatic charging of ice at 15--160 K. Physical Review B 85, 035424.

\bibitem[Shirley et al.(2010)]{shir10} Shirley, J.H., Dalton III, J.B., Prockter, L.M., Kamp, L.W., 2010.
Europa's ridged plains and smooth low albedo plains: Distinctive compositions
and compositional gradients at the leading side–trailing side boundary. Icarus, 210, 358--384.

\bibitem[Spencer et al.(1995)]{spe95}Spencer, J.R., Calvin, W.M., Person, M.J., 1995.
Charge-coupled device spectra of the Galilean satellites: 
Molecular oxygen on Ganymede. Journal of Geophysical Research 100, 19049.

\bibitem[Stillman et al.(2013)]{sti13}Stillman, D.E., MacGregor, J.A., Grimm, R.E., 2013.
The role of acids in electrical conduction through ice. Journal of Geophysical Research 118, 1--16.

\bibitem[Sugiyama et al.(2010)]{sug10}Sugiyama, S., Enomoto, H., Fujita, S., Fukui, K.,
Nakazawa, F., Holmlund, P., 2010. Dielectric permittivity of snow measured along the route traversed
in the Japanese-Swedish Antarctic Expedition 2007/08. Annals of Glaciology 51, 55.

\bibitem[Wang et al.(2008)]{wan08}Wang, H.F., Bell, R.C., Iedema, M.J., Schenter, G.K., Wu, K.,
and Cowin, J.P., 2008. Pyroelectricity of water ice. Journal of Physical Chemistry B, 112, 6379--6389.

\bibitem[Wieser et al.(2016)]{wie16}Wieser, M., Futaana, Y., Barabash, S., Wurz, P., 2016. 
Emission of energetic neutral atoms from ice under Ganymede surface like conditions. Icarus 269, 91--97.

\end{thebibliography}
\end{document}